\documentclass[%
 reprint,
 amsmath,amssymb,
 aps, showkeys,
]{revtex4-2}

\usepackage{graphicx}% Include figure files
\usepackage{dcolumn}% Align table columns on decimal point
\usepackage{bm}% bold math
\usepackage{xcolor}
\begin{document}

\preprint{APS/123-QED}

\title{Diffusion of fast and slow excitons with an exchange in quasi-two-dimensional systems}% Force line breaks with \\

\author{Oluwafemi P. Adejumobi}
 \affiliation{Skolkovo Institute of Science and Technology, Bolshoy Boulevard 30, Moscow, 121205, Russia.}
\author{Vladimir N. Mantsevich}%
\affiliation{%
 Chair of Semiconductors and Cryoelectronics, Physics department, Lomonosov Moscow State University, 119991 Moscow, Russia
}
\author{Vladimir V. Palyulin}
\email{v.palyulin@gmail.com}
\affiliation{Skolkovo Institute of Science and Technology, Bolshoy Boulevard 30, Moscow, 121205, Russia.
}

\date{\today}% It is always \today, today,

\begin{abstract}
By means of analytical calculations and numerical simulations we study the diffusion properties in quasi-two-dimensional structures with two exciton subsystems with an exchange between them. The experimental realisation is possible in systems where fast and slow exciton subsystems appear. For substantially different diffusion coefficients of the species the negative diffusion can be observed, if one measures the transport properties of only a single subsystem, just as was obtained in experimental studies for quasi-two-dimensional semiconductor systems. The initial transition from a fast subsystem to a slow one results in a delayed release of fast excitons in the area close to the original excitation spot. Hence, the signal from the fast excitons alone includes the delayed replenishment from the transition of slow quasi-particles. This is seen as the narrowing of the exciton density profile or decrease of mean-squared displacement which is then interpreted as a negative diffusion. The average diffusion coefficients for the combined population are analytically expressed through the diffusion coefficients of fast and slow excitons. Simple analytical expressions for effective diffusion coefficients obtained in limiting cases are of interest both for theoretical and experimental analysis of not only the exciton transport, but also for a variety of systems, where fast and slow moving subsystems are present.
\end{abstract}

 %Use showkeys class option if keyword
\maketitle

\section{\label{sec:level1}Introduction}

The light absorption in a semiconductor leads to an excitation of electrons from the valence band to the conduction band and creates quasiparticles called holes in the valence band \cite{ivchenko2005}. Due to Coulomb interaction electrons and holes form electron-hole pairs, known as excitons \cite{frenkel1931}. Excitons were experimentally observed in a wide variety of systems. Among them are inorganic and organic molecular semiconductors \cite{pope1999,colby2010,berghuis2021}, polymers \cite{hadziioannou2000,bolinger2011,kim2007}, colloidal quantum dots and nanoplatelets \cite{nanoplateletschemrev2023, klimov2000,ithurria2011,smirnov2019,smirnov2019_1,Rabouw2016,Shornikova2018,Olutas2015,Brumberg2019,Biadala2014,Meerbach2019}, semiconductor nanostructures (quantum wells and quantum wires) \cite{takagahara2003,scholes2006,grim2014}, perovskites \cite{Ishihara1989,Mitzi1994,Becker2018,Belykh2019,Ziegler2020,Magdaleno2021,Seitz2020} and transition metal dichalcogenides (TMDs) \cite{Mak2010,Splendiani2010,Chernikov2014,Wang2018,Cordovilla2018,chernikov2023}. 

In quasi-two-dimensional semiconductors such as quantum wells or transition metal dichalcogenides mono- and bi-layers excitons could move around the whole area of the material. The exciton transport mostly defines optoelectronic properties of semiconductors. The latter are utilised in a variety of applications such as novel lasers \cite{klimov2000}, photodetectors \cite{Konstantatos2012}, light-emitting diods \cite{Baugher2014,Xing2018} and solar cells \cite{Yang2017,Smith2014}.  This effects in the growing interest for the transport properties of excitons in semiconductor materials \cite{Yuan2017,Cadiz2018,Kulig2018,Bardeen2014,Dong2015,Kurilovich2020,Kurilovich2022, Kurilovich2023,Glazov2019}. Experimental studies of exciton transport in 2D materials show diffusive propagation across hundreds of nanometers at ambient conditions \cite{Kumar2014,Yuan2017}. Among the most important parameters describing exciton propagation are the diffusion length and the diffusion coefficient. Experimentally, these parameters could be obtained by detecting photoluminescence absorption dynamics as a function of pump intensity \cite{Rabouw2016,Shaw2008,Stevens2001}, spatially and time-resolved microphotoluminescence  \cite{Kulig2018,Akselrod2014,Han2018,Ginsberg2020} and comparison of photoluminescence signals for semiconductors both with quenching sites being present and absent \cite{Markov2005,Mikhnenko2008,Kholmicheva2015}. One of the most promising recent methods for exciton propagation analysis is the application of strain to the 2D systems that helps to produce exciton currents \cite{Shahnazaryan2019,Dirnberger2021,Datta2022}. 

Theoretical analysis of exciton transport is usually based on well-known approaches to transport phenomena in semiconductor systems \cite{Shklovskii1984,Gantmakher1987} with specifics of excitons being taken into account. This includes the presence of disorder, inter-particle interactions (exciton-exciton interactions, exciton-phonon interactions \cite{Smirnov_2019,Smirnov2021,Kajino2021,Hoshi2017}) and the role of external potentials. The most widely used models for exciton transport analysis are semiclassical models based on Boltzmann kinetic equation \cite{Zipfel2020,Choi2023}. Semiclassical treatment can be extended to take into account quantum effects by using the approach based on the diagram technique \cite{Glazov2022}. The alternatives are based on microscopic considerations and include tight-binding models \cite{Kenkre1983,Heijs2005} or kinetic Monte Carlo calculations \cite{Akselrod2014_1,Miyazaki2012}. Another fruitful approach considers excitons as classical particles which perform a random diffusive motion. It started from calculation of probabilities of photoluminescence versus absorption \cite{rosenstock1961random,levinson1962emission, rosenstock1974random,den1983trapping} and first-passage times to exciton traps \cite{montroll1969random,balagurov1974random} and continued with the use of diffusion equations directly showing an important role of traps and static disorder \cite{vlaming2013,Kurilovich2020, kurilovich2021trapping, Kurilovich2022, Kurilovich2023,Lunt2009,Lee2015,Glazov2019}.

Recently, experimental investigations of exciton transport revealed that excitons often split into two sub-populations, slowly and rapidly diffusing ones or trapped and free ones. An exchange between  two subpopulations happens leading for such non-obvious phenomenon as non-linear or negative effective diffusivity \cite{zhao2003spatiotemporal,Kurilovich2023,Wietek2024,berghuis2021,beret2023nonlinear,bornschlegl2024dark}. Multi-component systems with different diffusion coefficients for each subsystem typically demonstrate a crossover in the transport characteristics, as slow and fast components dominate the exciton kinetics at various time intervals. For instance, fast and slow exciton subsystems can be formed by bright and dark excitons \cite{berghuis2021,Rosati2020,Rosati2021,bornschlegl2024dark}. Originally, a laser pulse excites bright excitons. Some of them recombine and emit photons directly while others transfer into dark excitons. The dark excitons in turn have a relatively high diffusion coefficient right after the transition from the bright state but then thermalise into a state with a much lower diffusion coefficient. In tetracene crystals (organic semiconductors) the initial laser excitation results in radiative excitonic singlets which further split into dark triplets \cite{berghuis2021}. The latter in turn could fuse into delayed singlets as a binary reaction. Similar singlet-triplet exchange was found to produce the effective negative diffusion in monolayers of WSe$_2$ as well \cite{beret2023nonlinear}. Another regime, when two sub-systems are present, appears when both excitons and electron-hole
plasma are present \cite{Wietek2024}. Co-existence of excitons and electron-hole plasma is due to the rather strong exciton-exciton interaction. Finally, two subsystems can be formed by free (mobile) and trapped (immobile) excitons \cite{Lin2016,Kurilovich2022} and can be theoretically described by the mobile-immobile model (MIM) \cite{Kurilovich2022,Kurilovich2023}. The first evidence of negative exciton diffusivity at the intermediate timescales in ZnSe \cite{zhao2003spatiotemporal} was attributed to acoustic-phonon scattering. The authors correctly state that the equivalence of the total exciton population and the measured PL profiles are not always valid while also claiming that the classical diffusion description is not valid in low-temperature regime. The proposed explanation, however, predicted oscillations (breathing) of PL profile which was not observed in experiment. Recent papers based on classical diffusion ideas provide explanation of negative diffusion due to non-linear interactions \cite{Wietek2024} or an interplay of free and trapped subpopulations \cite{Kurilovich2023}. 

In this paper we modify MIM approach to analyse the exciton transport for systems with fast and slow excitons both analytically and numerically. We present analytical expressions for the distribution widths and numbers of excitons in free and trapped states, analyse limiting regimes and demonstrate the appearance of negative exciton diffusion regime. The long-limit formula for the diffusion of excitons also provides a way of measuring of rate constants. Obtained results are of interest for theoretical and experimental analysis of both exciton transport and a variety of systems, where fast and slow moving subsystems are present.

The paper is organised as follows. In section II we describe the simulation and analytical models and show how negative diffusion appears. In Section III we show expressions and plots for the exciton population numbers as a function of time, with the number of the fast excitons being proportional to the luminescence signal. In section IV we conclude and discuss the main results.

\section{Negative diffusion in the system with fast and slow excitons}

\begin{figure*}
    \includegraphics[width=1.0\linewidth]{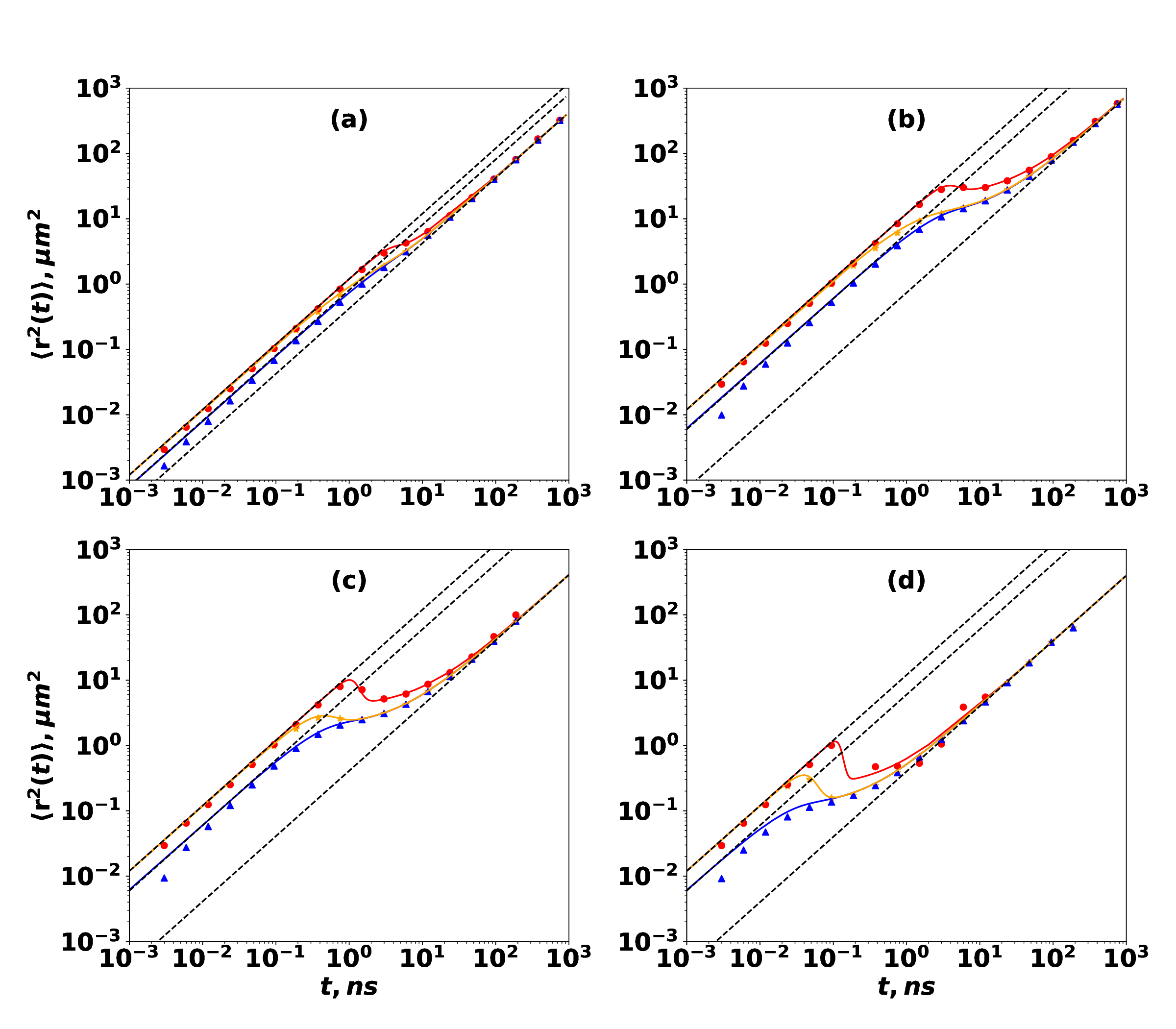}
    
    \caption{Average squared width of exciton distribution for fast, slow and all excitons. The continuous lines correspond to the fast (red), the slow (blue) and the overall (orange) subsystems of excitons and are plotted from expression (\ref{MSDnorm}). The black dashed straight lines with the slope 1 are plotted according to Eqs. (\ref{analytshorttimeWidthfree})-(\ref{analytshorttimeWidthtotal}) for the short times' limiting case and for the long times' case from Eq. (\ref{Eq:WidthfreeLongAnalyt}). Symbols of the corresponding colours are the simulation results. (a) $\alpha=0,D_A=0.3\,\mu m^2/\mathrm{ns},D_B=0.1\,\mu m^2/\mathrm{ns},\lambda_A=1\, \mathrm{ns^{-1}},\lambda_B=0.03\, \mathrm{ns^{-1}},N_0=10^5$, (b)  $\alpha=0, D_A=3\,\mu m^2/\mathrm{ns}, D_B=0.1\,\mu m^2/\mathrm{ns},\lambda_A=1\, \mathrm{ns^{-1}},\lambda_B=0.03\, \mathrm{ns^{-1}},N_0=10^5$, (c) $\alpha=5\, \mathrm{ns^{-1}},D_A=3\,\mu m^2/\mathrm{ns},D_B=0.1\,\mu m^2/\mathrm{ns},\lambda_A=1\, \mathrm{ns^{-1}},\lambda_B=0.03\, \mathrm{ns^{-1}},N_0=5\cdot10^5$, (d) $\alpha=100\, \mathrm{ns^{-1}},D_A=3\,\mu m^2/\mathrm{ns}, D_B=0.1\,\mu m^2/\mathrm{ns}, \lambda_A=1\, \mathrm{ns^{-1}},\lambda_B=0.03\, \mathrm{ns^{-1}},N_0=2\cdot10^6$.}
    \label{Fig:MSD}
\end{figure*}
 
The main idea of both of our analytical and simulation approaches is in splitting the diffusing excitons into two subsystems. One of them describes the slowly diffusing particles and the other corresponds to the fast walkers. 

In the experiments one measures the local PL signal enabling the assessment of the exciton density distribution as well as the overall photoluminescence output as a function of time \cite{berghuis2021,deng2020long}. Due to the exchange between the slow and the fast excitons and diffusion the initial Gaussian shape of the exciton density will change its form and become non-Gaussian (although it could be close to the Gaussian). The density can be analysed in different ways \cite{berghuis2021,beret2023nonlinear,Rosati2020,Rosati2021}. For instance, one could plot the cross-section of exciton densities normalised by the maximum value of the distribution \cite{berghuis2021}. Alternatively one considers the full width at half maximum (FWHM) or the mean-squared width of the distribution normalised by the number of excitons as in Refs. \cite{Rosati2020,Rosati2021}. Both methods account for the drop of exciton numbers in time otherwise the normalisation happens over the constant initial number of excitons and ensemble averages such as MSD tend to zero.

\subsection{The simulation model}

Our simulation model is a random walk in the plane with lengths of steps defined by the corresponding diffusion coefficients and switching between the "fast" and the "slow" states with diffusion coefficients $D_A$ and $D_B$, correspondingly, similar in spirit to the model from Refs. \cite{Kurilovich2020,Kurilovich2022,Kurilovich2023,kurilovich2021trapping}. Thus, the trajectory will contain parts with small steps alternating with stretches with longer steps (see Appendix A for the sketch).  At every step the jumping directions are equiprobable in 4 different directions along two perpendicular axes. Some of the semiconductor materials such as TMDs could have a hexagonal rather than square symmetry of the underlying lattice. However, we have shown previously \cite{Kurilovich2022} that results in these types of problems self-average well and no difference between simulations for the square and the triangle lattice were found. Excitons could also recombine and, thus, emit a photon. We gather statistics from $N_0$ single particle simulations since no binary or higher order interactions or reactions between excitons are modelled here  \cite{Kurilovich2020,Kurilovich2022}. A walker starts at the origin and then the random walk trajectories are generated and analysed.

Each simulation step has the duration $t_{\mathrm{step}}=10^{-3}\,\mathrm{ns}$ and consists of a few actions:
    1. An exciton performs a step with a length according to its state calculated from $\Delta x_{A,B}=\sqrt{4 D_{A,B}t_\mathrm{step}}$; 2. Faster excitons recombine radiatively with the rate $\alpha$ (ordinary recombination). The number of these events could then be counted and after binning and normalisation could be used to produce a photoluminescence curve \cite{Rabouw2016}; 3. The particle chooses whether to change the fast/slow state according to Markovian transition rates $\lambda_A,\lambda_B$. The latter define transition probabilities as $p_{AB,BA}=\lambda_{A,B}t_\mathrm{step}$. 

\subsection{The analytical model}

De facto our simulation model is microscopic discretisation of a diffusion process which also could be represented by continuous diffusion-reaction equations as considered below. Our analytical model considers a two-state system with an exchange and recombination for one type of the species similar to the MIM model with temporary immobilisation of excitons by traps \cite{Kurilovich2020,Kurilovich2022,Kurilovich2023}. Previous model analysis \cite{Kurilovich2023} revealed that binary interactions between excitons enhance the negative diffusion visibility, however, the effect can be observed without them. We focus here on the interplay of the slow and fast diffusing particles and do not include high-order interactions. The exciton-exciton scattering can be neglected for experiments performed with low exciton density  \cite{zhao2003spatiotemporal}.

For the analytical description with the use of densities $n_A (\mathbf{r},t)$ and $n_B (\mathbf{r},t)$ for excitons of types A and B, respectively. Likewise, the total density reads $n (\mathbf{r},t) = n_A (\mathbf{r},t) + n_B (\mathbf{r},t)$. Then the corresponding reaction-diffusion equations for the densities are
\begin{eqnarray}\label{n_A}
\frac{d n_A (\mathbf{r}, t)}{dt} && = D_A\Delta n_A(\mathbf{r},t) - \lambda_A n_A(\mathbf{r},t)  \\\nonumber && +\lambda_B n_B(\mathbf{r}, t) - \alpha  n_A(\mathbf{r},t) ,\\\label{n_B}
\frac{d n_B(\mathbf{r},t)}{dt} &&= D_B\Delta n_B(\mathbf{r},t) + \lambda_A  n_A(\mathbf{r},t)   \\\nonumber && -  \lambda_B n_B(\mathbf{r}, t).
\end{eqnarray}
In experiments right after a laser excitation all excitons are in the fast state. Hence we set the initial conditions as $n_A (\mathbf{r}, t=0)= N_0 \delta(r),$ and $n_B (\mathbf{r}, t=0)=0$, where $N_0$ is the number of produced particles. The boundary conditions are zero values for the densities and their first derivatives for $\vert\mathbf r\vert\rightarrow\infty$. The numbers of A and B excitons $N_{A,B}(t)$ can then be obtained by integration of densities over the plane,
\begin{eqnarray}\label{n_AB}
N_{A,B}(t) = \int n_{A,B} (\mathbf{r}, t) d\mathbf{r}.
\end{eqnarray}
Solving (\ref{n_A})-(\ref{n_B}) analytically in the Fourier-Laplace space, we obtain
\begin{eqnarray}
    &n_A (\mathbf{k}, s) = \frac{N_0 (s + \lambda_B + D_B k^2)}{\left(s + \alpha + \lambda_A + D_A k^2\right)\left(s + \lambda_B + D_B k^2\right) - \lambda_A \lambda_B},\label{typeAdensitysolution}\\
    &n_B (\mathbf{k}, s) =\frac{N_0 \lambda_A}{\left(s + \alpha + \lambda_A + D_A k^2\right)\left(s + \lambda_B + D_B k^2\right) - \lambda_A \lambda_B}, \label{typeBdensitysolution}
\end{eqnarray}
The image of the total concentration of all excitons $n(\mathbf{r},t)$ is then
\begin{equation}
    n(\mathbf{k}, s) = \frac{N_0 [s + \lambda_A + \lambda_B+ D_B k^2 ]}{\left(s + \alpha + \lambda_A + D_A k^2\right)\left(s + \lambda_B + D_B k^2\right) - \lambda_A \lambda_B}.
\end{equation}
The width of the distribution or mean-squared displacement for both subsystems can be found from the raw second moment of densities $I_{A,B}(t)$ and needs to be normalised by the corresponding exciton numbers (see also the discussion in Refs. \cite{Kurilovich2023},\cite{Rosati2021}). This produces the quantity equivalent to mean-squared displacement (MSD) and is given by:
\begin{eqnarray} \label{MSDnorm}
&&\langle \vert r(t) \vert^2\rangle_{A,B} =\frac{I_{A,B}(t)}{N_{A,B}(t)} \label{Eq:WidthDef} , \\
\text{where} \quad &&I_{A,B}(t) = \int \vert \mathbf r\vert^2 n_{A,B}(\mathbf{r},t)d\mathbf{r}. \nonumber
\end{eqnarray}

Normalisation is necessary since the exciton numbers change and eventually tend to zero due to the recombination. The Laplace images of $I_{A,B}(t)$ are

\begin{eqnarray}\label{Eq:WidthNonFree}
I_{A}(s) &&= \frac{4N_0 [D_B \lambda_A \lambda_B + D_A(s+\lambda_B)^2]}{(s^2+\alpha\lambda_B+ s(\alpha + \lambda_A + \lambda_B))^2},\\\label{Eq:WidthNonTrapped}
I_{B}(s) &&= \frac{4N_0 \lambda_A[D_B ( s + \alpha +\lambda_A) + D_A(s+\lambda_B)]}{(s^2+\alpha\lambda_B+ s(\alpha + \lambda_A + \lambda_B))^2},\\
I(s) &&= I_A(s)+I_B(s).\label{Eq:WidthNonTotal}
\end{eqnarray}
The inverse Laplace transform of these expressions can be performed but the result is rather cumbersome and is shown in Appendix \ref{App:Widthanalyt}.

While in experiment one cannot track every exciton it is possible to do that in simulation studies, i.e. one could plot mean-squared displacement for a single particle including information about its state. In Fig. \ref{Fig:MSD} the continuous lines show the mean-squared width obtained from Eqs. (\ref{n_A})-(\ref{n_B}) and the symbols correspond to the simulated mean-squared displacement of a single exciton averaged over the ensemble of A, B or all excitons. The MSD is normalised by the number of surviving excitons as is done in experiments \cite{Rosati2020,Rosati2021}. The parameters chosen are within the experimentally measured range \cite{chernikov2023,Cadiz2018,Zipfel2020,Wietek2024}. We show four typical cases for the mean-squared width of the distribution of excitons. Initially the MSD for all particle subsystems is linear, i.e. excitons perform a simple random walk. For a small difference between diffusion coefficients the MSD values continue to grow monotonously for all excitons (Fig. \ref{Fig:MSD}a). However, if the timescale of being in the slow state is substantially longer than for the fast state while the difference in diffusion coefficients is substantial, then the fast excitons after a while are replenished with the formerly slow excitons caught in the slow state close to the origin. This leads to the appearance of the negative slope followed by a plateau-like region (Fig. \ref{Fig:MSD}b). This phenomenon is essentially transitional and the diffusion again becomes Brownian at long times and the slope can be computed analytically.
If we add linear recombination, i.e. $\alpha>0$ (Figs. \ref{Fig:MSD}c,d), the visible effect of the negative diffusion gets enhanced (blue curve) and the plateau extends. Furthermore, for very high recombination rates even the overall number of excitons could show a slight negative MSD slope (the orange curve in Fig. \ref{Fig:MSD}d).

The short time and the long time limits can be expressed with tractable simple expressions from exact solutions from Appendix \ref{App:Widthanalyt} and are shown as dashed lines in Fig. \ref{Fig:MSD}. At short times $t\rightarrow0$,  
\begin{eqnarray}\label{analytshorttimeWidthfree}
&&\langle \vert r(t) \vert^2\rangle_{A}\simeq 4D_A t -\frac{2 \lambda_A \lambda_B }{3} (D_A - D_B) t^3+...\\ \nonumber
&&\langle \vert r(t) \vert^2\rangle_{B}\simeq2(D_A + D_B) t\\ \nonumber
&&-\frac{1}{3}((D_A - D_B)(\alpha + \lambda_A - \lambda_B))t^2+...,\\\label{analytshorttimeWidthtrapped}
\\\nonumber
&&\langle \vert r(t) \vert^2\rangle\simeq4D_At-2\lambda_A(D_A - D_B)t^2+... .\\\label{analytshorttimeWidthtotal}
\end{eqnarray}
For $D_B=0$, i.e. when slow excitons become immobile these equations reduce to the formulas of MIM model, Eqs. (14)-(16) from Ref. \cite{Kurilovich2023} (unfortunately, the correction terms to the linear dependence in the reference have a missing diffusion coefficient $D$). Thus, at short times, for fast excitons the normal diffusion (the slope equals to unity) proceeds with the coefficient corresponding to the regular jumps between the neighboring lattice sites. However, the spread of the population of slow excitons is already controlled by the average of diffusion coefficients for the fast and the slow states.  

At long times $t\rightarrow\infty$, the MSD of all populations of excitons converges to the same behaviour (for the derivation see Appendix \ref{App:Widthanalyt}),
\begin{equation}\label{Eq:WidthfreeLongAnalyt}
    \langle \vert r(t) \vert^2\rangle_{A}=\langle \vert r(t) \vert^2\rangle_{B}=\langle \vert r(t) \vert^2\rangle=4C_1t,
\end{equation}
where $C_1$ is a constant (an effective long time diffusion coefficient) which depends on parameters $\alpha,\lambda_A,\lambda_B,D_A,D_B$ and is defined in Appendix \ref{App:Widthanalyt}. For $\alpha=0$ expressions are further simplified and read
\begin{equation}
    \langle \vert r(t) \vert^2\rangle_{A}=\langle \vert r(t) \vert^2\rangle_{B}=\langle \vert r(t) \vert^2\rangle=\frac{4 (D_A \lambda_B+D_B\lambda_A) t}{\lambda_A+\lambda_B},
\end{equation}
i.e. the long term diffusion coefficient is defined by the transition rates as well as the random walk in between. For $D_B=0$ one again obtains the corresponding result from \cite{Kurilovich2023}. For $\alpha>0$ expressions for normalised mean-square widths also coincide and $C_1$ consists of two terms,
\begin{eqnarray}\nonumber
    C_1 =\frac{1}{2}(D_A + D_B)-\frac{1}{2\sqrt{A}} (D_A - D_B) (\alpha + \lambda_A - \lambda_B),
\end{eqnarray}
where the shorthand notation $A$ is defined in Appendix \ref{App:Widthanalyt}. In the limit $D_B=0$ the result Eq. (32) from the Ref. \cite{Kurilovich2023} is reproduced. For large values of $\alpha$ the last expression reduces to $C_1=D_B$.   

Hence, the interplay of the diffusion properties enters as a difference $(D_A - D_B)$. When the difference vanishes the MSD recovers a normal linear dependence. The increase of that difference shifts the long time limit extends the plateau region. The growth of $\alpha$ also pushes the spread between the short and the long time limits, thus, elongates the plateau.

An important conclusion from the analysis of the time limits is in the usefulness of the long time limit for the measurements of transition rates if diffusion coefficients and $\alpha$ are known. 
\begin{figure}
    \includegraphics[width=\columnwidth]{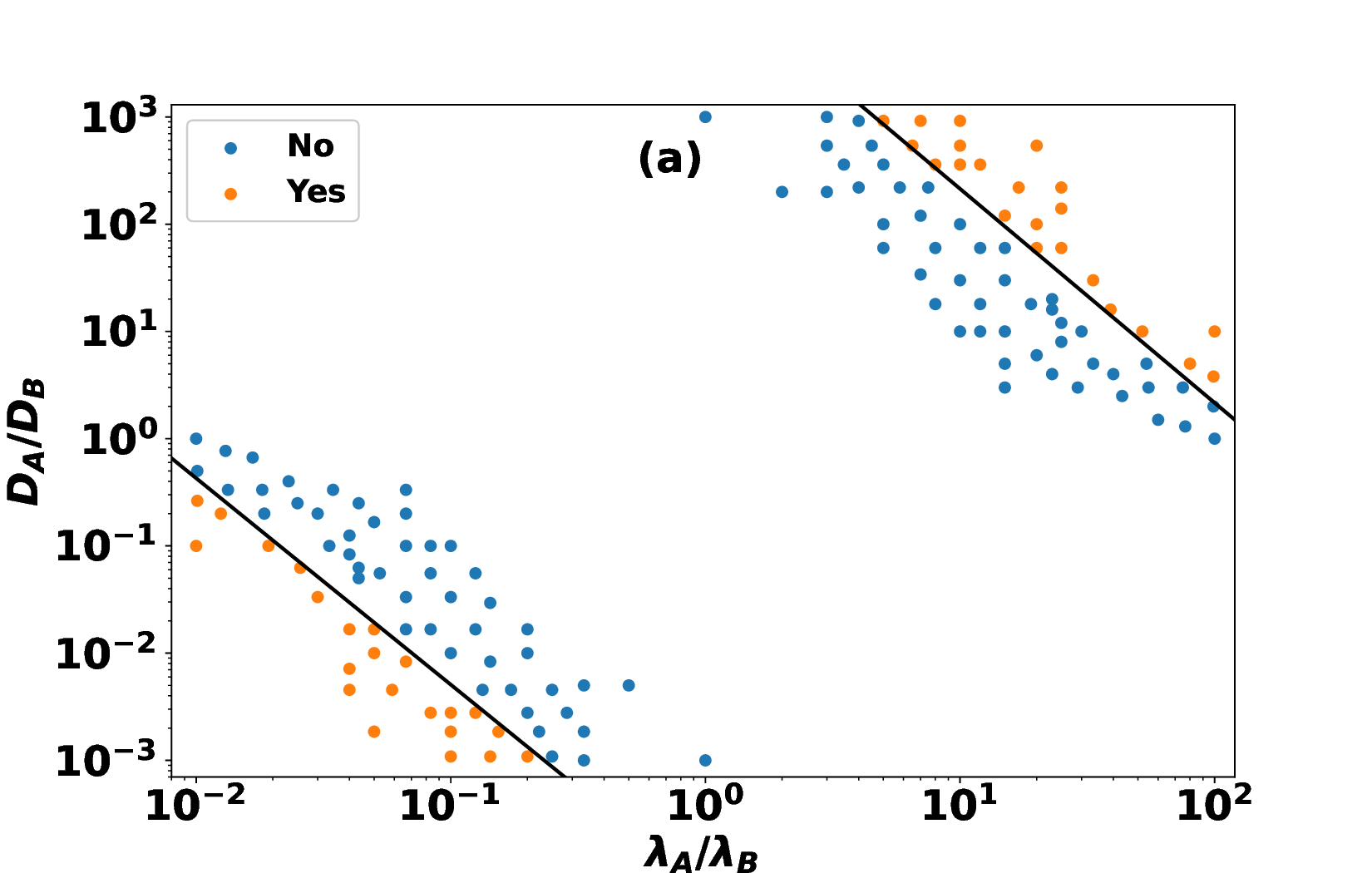}
    \includegraphics[width=\columnwidth]{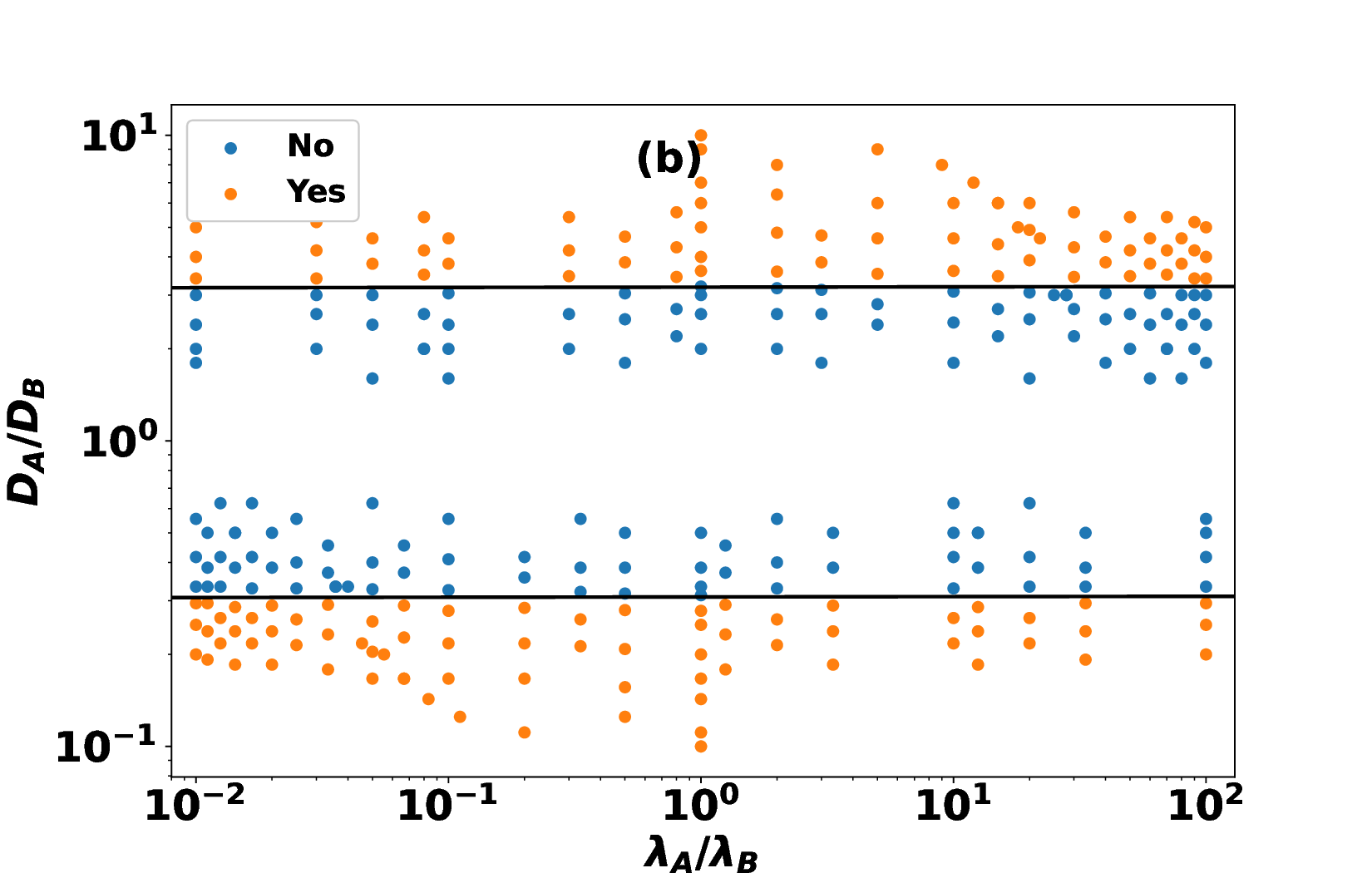}
    \caption{The diagram showing the parameter ranges with/without negative diffusion. $D_B=1\,\mu m^2/\mathrm{ns},\lambda_B=1\, \mathrm{ns}^{-1}$. In the upper diagram $\alpha=0$, while in the lower one $\alpha=5\, \mathrm{ns}^{-1}$. The orange dots highlight the region with negative diffusion. The blue dots correspond to its absence. The black lines are drawn to guide the eye rather than a calculated boundaries.}
    \label{Fig:Phase diagram}
\end{figure}

In Fig. \ref{Fig:Phase diagram} we show the how the diffusion coefficient ratios $D_A/D_B$ and the ratio of transition rates $\lambda_A/\lambda_B$ affect the appearance of the negative diffusion which was checked by plotting MSD curves with the analytical solution (\ref{Eq:WidthDef}) for subsystem A. One can see that in order to get a negative diffusion for $\alpha=0$ one needs more extreme ratios. For relatively large recombination rates such as $\alpha=5\,\mathrm{ns}^{-1}$ the transition rates cease to play a role and one needs just a substantial difference of diffusion coefficients of two exciton types. 

\begin{figure*}
    \includegraphics[width=\linewidth]{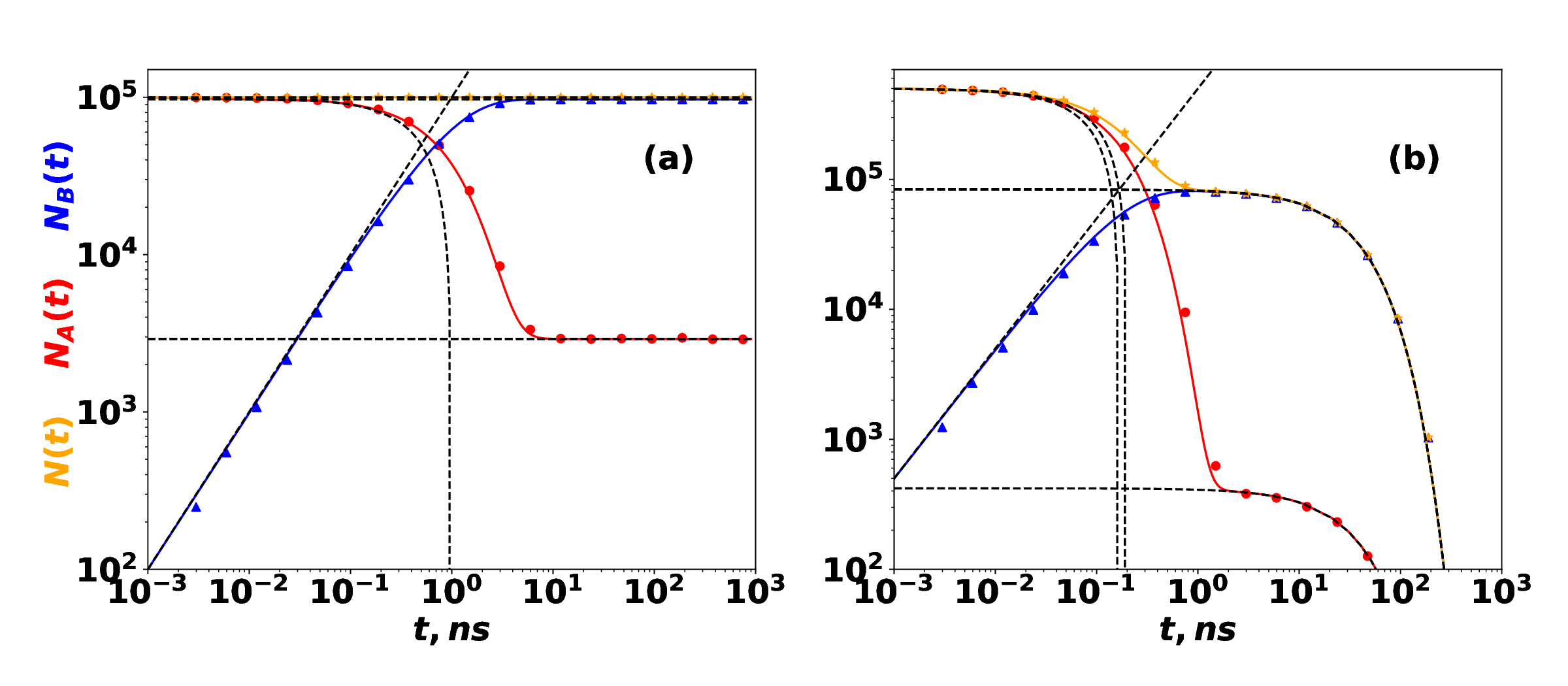}
    \caption{Theoretical and simulation results for the numbers of free, trapped and all excitons for the linear recombination and without it. The continuous lines correspond to free (red), trapped (blue) and overall (orange) numbers of excitons and are plotted from analytical formulae (\ref{typeAnumber})-(\ref{alltype_number}). The black dotted lines correspond to the short and long time limiting behaviour according to Eqs. (\ref{analytshorttimeNA})-(\ref{analytshorttimeNtotal}) and (\ref{NumbersLimiting}), correspondingly. (a)  $\alpha=0, D_A=3\,\mu m^2/\mathrm{ns}, D_B=0.1\,\mu m^2/\mathrm{ns}, \lambda_A=1\, \mathrm{ns^{-1}},\lambda_B=0.03\, \mathrm{ns^{-1}}$, $N_0=10^5$, (b) $\alpha=5\, \mathrm{ns^{-1}}, D_A=3\,\mu m^2/\mathrm{ns}, D_B=0.1\,\mu m^2/\mathrm{ns}, \lambda_A=1\, \mathrm{ns^{-1}},\lambda_B=0.03\, \mathrm{ns^{-1}}$, $N_0=5\cdot 10^5$.}
    \label{Fig:N}
\end{figure*}

\section{NUMBER OF EXCITONS AS A FUNCTION
OF TIME}

The overall number of the corresponding subspecies of excitons can be obtained from Eqs.(\ref{typeAdensitysolution})-(\ref{typeBdensitysolution}) just by considering $\mathbf{k}=\mathbf{0}$ since 
\begin{eqnarray}
\tilde N_{A,B}(s) = \int \tilde n_{A,B}(\mathbf{r}, s)d\mathbf{r} = \hat{\tilde{n}}_{A,B}(\mathbf{k}=\mathbf{0}, s),
\end{eqnarray}
i.e.
\begin{eqnarray}
    &\tilde N_A (s) = \frac{N_0 (s + \lambda_B )}{\left(s + \lambda_B\right) \left(s + \alpha + \lambda_A \right) - \lambda_A \lambda_B},\label{typeAnumbersolution}\\
    &\tilde N_B (s) = \frac{N_0 \lambda_A }{\left(s + \lambda_B\right) \left(s + \alpha + \lambda_A \right) - \lambda_A \lambda_B}.
    \label{typeBnumbersolution}
\end{eqnarray}
It is possible to analytically invert Eqs. (\ref{typeAnumbersolution})-(\ref{typeBnumbersolution}) from Laplace space to real time  (the full analytical expressions can be found in the Appendix \ref{App:Nanalyt}, (\ref{typeAnumber})-(\ref{alltype_number})). While the general expression looks more cumbersome, the asymptotics look simpler. For $t\rightarrow0$, 
\begin{eqnarray}\label{analytshorttimeNA}
&&N_A(t)\simeq N_0(1-(\alpha+\lambda_A)t+...),\\ \label{analytshorttimeNB} 
&&N_B(t)\simeq N_0\lambda_A t,\\\label{analytshorttimeNtotal} &&N(t)=N_0(1-\alpha t).
\end{eqnarray}
Expressions in the limit  $t\rightarrow\infty$ can be worked out analytically by leaving the leading term in the corresponding Eqs. (\ref{typeAnumber})-(\ref{alltype_number}) which result in the formulae (\ref{NumbersLimiting}) shown in Appendix \ref{App:Nanalyt}. One can see that the long-term decay is exponential with the same exponent for all exciton types.

The results depicted in Fig. \ref{Fig:N} demonstrate two distinct scenarios. First, in the absence of recombination, the system exhibits an exponential decay in the number of fast excitons (red curve/symbols) due to their transition to slow excitons (blue curve/symbols), while the number of slow excitons exhibits linear growth until an equilibrium is reached between two states. The dashed lines show analytical short and long time limits. Second, in the presence of recombination (Fig. \ref{Fig:N}b), a similar exponential decay is observed for fast excitons (red curve/symbols) at short times, attributed to their transition to slow excitons (blue curve/symbols) coupled with recombination. Meanwhile, the number of slow excitons increases linearly, resembling the behaviour in the absence of recombination. However, a significant deviation from Fig. \ref{Fig:N}a occurs at long times, where the numbers of all exciton subsystems experience an exponential decay. Again the analytical limits Eqs. (\ref{analytshorttimeNA})-(\ref{analytshorttimeNtotal}) and (\ref{typeAnumber})-(\ref{alltype_number}) are shown as 
dashed lines and reveal the perfect fit. Under the assumption of only linear recombination the overall luminescence intensity will be proportional to the number of fast excitons, i.e. the red curve in Fig. \ref{Fig:N}b.

\section{Conclusions and Discussion}

The exciton diffusion in semiconductors 
often shows deviation from a classical Brownian diffusive motion \cite{Rabouw2016,Seitz2020,Rosati2020,Rosati2021,berghuis2021,Ziegler2020,Akselrod2014,Kulig2018}, thus, extending an abundant universe of anomalous diffusion \cite{metzler2014anomalous,sokolov2012models}. The reasons for the deviation are plenty. To name the few, excitons can be trapped and released, coexist in singlet/triplet or bright/dark states. They can interact with phonons and with each other leading to annihilation or recombination. We show here that despite a seeming non-linearity and complexity of the underlying physics simple Markovian models can be useful for the explanation of some of the effects, e.g. negative diffusion at intermediate scales. While the negative diffusion could appear due to non-linear (for instance, binary) interactions we propose a mechanism based on an interplay of two subsystems of excitons with substantially different diffusion coefficients and an exchange. The negative slope and the plateau occur due to the initial trapping of fast particles close too the point of generation and their delayed release. Thus, it is similar in spirit to the negative diffusion explanation with MIM \cite{Kurilovich2023}. The two-state mobile-mobile model also produces MSD plateaus for some parameters which is a rare instance for anomalous diffusion models. Among the other systems one could mention MSD of non-normalisable quasi-equilibrium states under fractional dynamics \cite{Eli2023} or probe particle tracking experiments in micellar solutions with subdiffusive dynamics \cite{bellour2002brownian,galvan2008diffusing,jeon2013anomalous} as well as in the cytosol of bacteria, yeast, and human cells \cite{aaberg2021glass,corci2023extending}.

Our approach is an extension or variation of mobile-immobile models \cite{Kurilovich2022,Kurilovich2023,doerries2022apparent,doerries2023emergent} and it shows an anomalous diffusion as also happens in systems with two Markovian hopping-trap mechanisms \cite{vitali2022anomalous}. However, the reaction-diffusion kinetics with an exchange is wider in scope and includes more complex formulations, for instance, such as equations with power-law residence times \cite{fedotov2011non,han2021anomalous,korabel2023non} or non-equilibrium reaction-diffusion systems with binary reactions \cite{goppel2016efficiency}.

Contrary to our previous mobile-immobile models for the excitons diffusing in the field of traps \cite{Kurilovich2022,Kurilovich2023} we get a perfect match between the random walk based simulation model and the analytical model. In the case of simulation of two-state model with exchange between two mobile states in this paper a frozen disorder does not appear, the system self-averages and the behaviour converges to the solutions of diffusion-reaction equations. In the case of the MIM simulations \cite{Kurilovich2022} the initialisation of the trap distribution sets the frozen disorder which affects the intermediate times properties quantitatively although qualitatively the behaviour stays the same.

\appendix
\section{Sketch of a trajectory in simulation model}

\begin{figure}
    \includegraphics[width=\columnwidth]{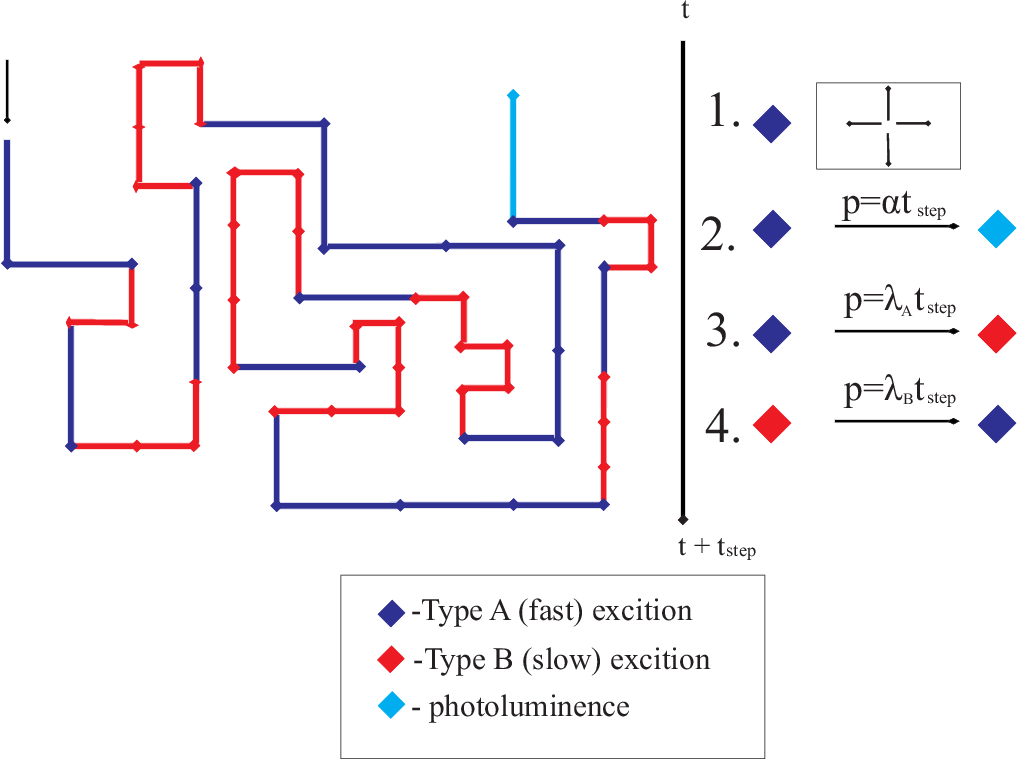}
    \caption{Sketch of a trajectory in the simulation model}
    \label{Fig:sketch}
\end{figure}

In Figure \ref{Fig:sketch} a short piece of a sample trajectory is depicted. At every timestep the exciton picks the direction out of 4 perpendicular ones, could recombine and make a short or a long step according to its diffusion coefficient, and after that it could change the state.
\section{Analytical formulas for the mean-squared displacement of excitons}\label{App:Widthanalyt}

Considering all the subsets derived from Eq. \ref{Eq:WidthDef}, the mean-square displacement expression may look complex but here is a simplified form by getting the expression of each integral $I_A(t), I_B(t), I(t)$ and dividing by the corresponding number of excitons given in Eq. \ref{Eq:WidthDef}.

Denoting $A = -4\alpha\lambda_B + (\alpha + \lambda_A+\lambda_B)^2$
one gets
\begin{align}
    I_A(t) &= \frac{2N_0 e^{-\frac{1}{2} t \left(\alpha + \lambda_A + \lambda_B\right)} }{A^{3/2}} \left[ e^{-\frac{1}{2} t \sqrt{A}} \left(2 D_B \lambda_A \lambda_B (2 + \sqrt A t) \right.\right.\nonumber \\
&+ D_A \left(\alpha (2 \sqrt A + 3 \alpha) \lambda_A t + \lambda_A^3 t \right.\nonumber \\&+ (\alpha - \lambda_B)^2 (\sqrt A + \alpha - \lambda_B) t - 
    \lambda_A \lambda_B^2 t \nonumber \\
&\left.\left.+\lambda_A^2 (\sqrt A + 3 \alpha + \lambda_B) t - 
    2 \lambda_A \lambda_B (2 + \alpha t)\right)\right) \nonumber \\
    &+ e^{\frac{1}{2} t \sqrt{A}}  \left(2 D_B \lambda_A \lambda_B (-2 + \sqrt A t) \right.\nonumber \\
    &+D_A \left(-\lambda_A^3 t + 
    \lambda_A^2 (\sqrt A - 3 \alpha - \lambda_B) t \right.\nonumber \\
&+ (\alpha - \lambda_B)^2 (\sqrt A - \alpha + \lambda_B) t \nonumber \\
&\left.\left.+ \lambda_A ((2 \sqrt A - 3 \alpha) \alpha t + \lambda_B^2 t + 2 \lambda_B (2 + \alpha t)))\right) \right],
\end{align}

\begin{align}
    I_B(t) &= \frac{2 N_0\lambda_A e^{-\frac{1}{2} t \left(\alpha + \lambda_A + \lambda_B\right)}}{A^{3/2}} \left[ 
    e^{-\frac{1}{2} t\sqrt{A}} \left(-2(\alpha+\lambda_A-\lambda_B)\right.\right.\nonumber\\&\times(D_A - D_B)+t \left(-\alpha^2 (D_A + D_B) \right.\nonumber\\&- 
   D_A (\sqrt A \lambda_A + \lambda_A^2 - \sqrt A \lambda_B + 
      2 \lambda_A \lambda_B + \lambda_B^2) \nonumber\\&+ 
   D_B (-\lambda_A^2 + \lambda_A (\sqrt A - 2 \lambda_B) - 
      \lambda_B (\sqrt A + \lambda_B)) \nonumber\\&+ 
   \alpha \left(-D_A (\sqrt A + 2 \lambda_A - 2 \lambda_B) \right.\nonumber\\&\left.\left.+ 
      D_B (\sqrt A - 2 \lambda_A + 2 \lambda_B))\right)\right) \nonumber\\ 
    &+e^{\frac{1}{2} t\sqrt{A}}\left(2(D_A - D_B) (\alpha + \lambda_A - \lambda_B) \right.\nonumber\\ 
    &+ t\left(-\sqrt A (D_A - D_B) (\alpha + \lambda_A - \lambda_B)\right.\nonumber\\ 
    & \left.\left.\left.+ (D_A + D_B) \left(\alpha^2 + 
    2 \alpha (\lambda_A - \lambda_B) + (\lambda_A + \lambda_B)^2\right) \right)\right)
    \right],\\
    I(t) &= I_A(t) + I_B(t) .
\end{align}
Using expressions above, coupled with the results from Appendix \ref{App:Nanalyt} and Eq. (\ref{Eq:WidthDef}), one can derive the
limiting cases for the evolution of the mean squared width for the exciton position distributions. At short times the results can be found
in Eqs. (\ref{analytshorttimeWidthfree})-(\ref{analytshorttimeWidthtotal}).

In the long time limit, $t \rightarrow \infty$, it leads to the following expression for the mean-squared width of the distributions of fast, slow and all excitons,
\begin{equation}
    \langle r^2(t) \rangle_A = \langle r^2(t) \rangle_B = \langle r^2(t) \rangle = 4 C_1 t,
\end{equation}
where
\begin{equation}
    C_1 =\frac{1}{2}(D_A + D_B)-\frac{1}{2\sqrt{A}} (D_A - D_B) (\alpha + \lambda_A - \lambda_B).
\end{equation}

\section{Analytical formulas for the numbers of excitons}\label{App:Nanalyt}
Equations (\ref{typeAnumbersolution}) - (\ref{typeBnumbersolution}) can be analytically inverted to real time from Laplace space, to get,

\begin{align}
    &N_A (t) = \frac{N_0 (\alpha + \lambda_A - \lambda_B + \sqrt{A})}{2 \sqrt{A}} e^{-\frac{1}{2}t\left( \alpha + \lambda_A + \lambda_B + \sqrt{A}\right)} \nonumber \\ 
    &\quad \quad \quad + \frac{N_0 (- \alpha - \lambda_A + \lambda_B + \sqrt{A})}{2 \sqrt{A}} e^{-\frac{1}{2}t\left( \alpha + \lambda_A + \lambda_B - \sqrt{A}\right)}, \label{typeAnumber} \\
    &N_B (t) = \frac{N_0 \lambda_A }{\sqrt{A}}e^{-\frac{1}{2}t\left( \alpha + \lambda_A + \lambda_B \right)} \left[ e^{\frac{1}{2} \sqrt{A} t} - e^{-\frac{1}{2} \sqrt{A} t}  \right],\label{typeBnumber}\\
    &\quad N (t) = N_A (t) + N_B (t) \nonumber \\
    &\quad \quad \quad = \frac{N_0 (\alpha - \lambda_A - \lambda_B + \sqrt{A})}{2 \sqrt{A}} e^{-\frac{1}{2}t\left( \alpha + \lambda_A + \lambda_B + \sqrt{A}\right)} \nonumber \\ 
    &\quad \quad \quad  + \frac{N_0 (- \alpha + \lambda_A + \lambda_B + \sqrt{A})}{2 \sqrt{A}} e^{-\frac{1}{2}t\left( \alpha + \lambda_A + \lambda_B - \sqrt{A}\right)}, \label{alltype_number}
\end{align}
Naturally, when $\alpha = 0$, $N(t) = N_A (t)+N_B (t) = N_0 = const$.

But, from the expression above, for $t\rightarrow 0$, $N_A(t)\simeq N_0(1-(\alpha+\lambda_A)t+...), N_B(t)\simeq N_0\lambda_A t, N(t)\simeq N_0(1-\alpha t)$. For $\lambda_B t \gg 1 (s \rightarrow 0)$,
\begin{align}
    N_A (t) &= \frac{N_0 (- \alpha + \lambda_A + \lambda_B + \sqrt{A})}{2 \sqrt{A}} e^{-\frac{1}{2}t\left( \alpha + \lambda_A + \lambda_B - \sqrt{A}\right)}, \nonumber\\
    N_B (t) &= \frac{N_0 \lambda_A }{\sqrt{A}}e^{-\frac{1}{2}t\left( \alpha + \lambda_A + \lambda_B - \sqrt{A} \right)},\nonumber\\
    N (t) &= \frac{N_0 (- \alpha + \lambda_A + \lambda_B + \sqrt{A})}{2 \sqrt{A}} e^{-\frac{1}{2}t\left( \alpha + \lambda_A + \lambda_B - \sqrt{A}\right)}, \nonumber\\ \label{NumbersLimiting}
\end{align}

\acknowledgements

The authors would like to acknowledge the support of the Russian
Science Foundation project no. 24-12-00020.
\bibliography{Literature}% 

\end{document}